\documentclass[a4paper]{jpconf}
\usepackage{graphicx}
\begin{document}
\title{Delineating effects of tensor force on the density dependence of nuclear symmetry energy}

\author{Chang Xu$^{1,2}$, Ang Li$^{2,3}$ and Bao-An Li$^{2*,4}$}

\address{$^{1}$School of Physics, Nanjing University, Nanjing 210008, China}

\address{$^{2}$Department of Physics and Astronomy, Texas A$\&$M University-Commerce, Commerce, Texas 75429-3011, USA}

\address{$^{3}$Department of Physics and Institute of Theoretical Physics and Astrophysics, Xiamen University, Xiamen
361005, China}

\address{$^{4}$Department of Applied Physics, Xian Jiao Tong University, Xian 710049, China}

\ead{Chang Xu,Cxu@nju.edu.cn}
\ead{Ang Li,Liang@xmu.edu.cn}
\ead{$^*$Corresponding author:Bao-An Li,Bao-An.Li@Tamuc.edu}

\begin{abstract}
In this talk, we report results of our recent studies to delineate effects of the tensor force on the density dependence
of nuclear symmetry energy within phenomenological models.
The tensor force active in the isosinglet neutron-proton
interaction channel leads to appreciable depletion/population of
nucleons below/above the Fermi surface in the single-nucleon
momentum distribution in cold symmetric nuclear matter (SNM).
We found that as a consequence of the high momentum tail in SNM the kinetic part of the symmetry energy $E^{kin}_{sym}(\rho)$ is
significantly below the well-known Fermi gas model prediction of
approximately $12.5 (\rho/\rho_0)^{2/3}$. With about $15\%$ nucleons in the high momentum tail
as indicated by the recent experiments at J-Lab by the CLAS Collaboration,
the $E^{kin}_{sym}(\rho)$ is negligibly small. It even becomes negative
when more nucleons are in the high momentum tail in SNM. These features have recently been confirmed by
three independent studies based on the state-of-the-art microscopic nuclear many-body theories. In addition, we also
estimate the second-order tensor force contribution to the potential part of the symmetry energy. Implications of these findings
in extracting information about nuclear symmetry energy from nuclear reactions are discussed briefly.
\end{abstract}

\section{Introduction}

Nuclear symmetry energy $E_{sym}(\rho)$, which encodes the energy
related to neutron-proton asymmetry in the nuclear matter Equation
of State (EOS), is a vital ingredient in the theoretical
description of neutron stars and of the structure of neutron-rich
nuclei and reactions involving them. Since the density-dependence
of $E_{sym}(\rho)$ is still the most uncertain part of the EOS of
neutron-rich nucleonic matter especially at supra-saturation
densities, to better determine the $E_{sym}(\rho)$ has become a
major goal of both nuclear physics and astrophysics
\cite{ireview98,ibook01,dan,bar,li1,Sum94,Lat04,Ste05a,Xuli10a,Xuli10b,LWC05,tsa,Cen09,xia}.
While significant progress has been made recently in narrowing
down the symmetry energy near normal nuclear matter density $\rho_0$,
see, e.g., \cite{Nat10,Tsang11,Newton11,War11,Dut11,BALI11,Dong12,ts12,St12,Lat12}, much
more efforts are needed to pin down the $E_{sym}(\rho)$ at both
sub- and supra-saturation densities. Moreover, it is now broadly
recognized that essentially all of the constraints extracted from experimental data
are model dependent. Thus, to make further progress in
the field, it is imperative to identify clearly the key physics
ingredients determining the density dependence of nuclear symmetry
energy in each model \cite{Xuli10a,Far12}. Besides different techniques used in various nuclear
many-body theories, several ingredients, such as, the spin-isospin
dependence of the three-body force, tensor force induced high
momentum tail in the single-nucleon momentum distribution of
symmetric nuclear matter (SNM) and the associated
isospin-dependence of short-range two-nucleon correlations, the
isospin-dependence of nucleon pairing and clustering at low
densities, are particularly known to affect significantly the $E_{sym}(\rho)$.
Of course, these ingredients may be approximately equally important and interfere strongly at some
densities but individually dominate at other densities in models
where they are all considered. In reality, however, they are rarely
all taken into account simultaneously in a given model. Also,
among these ingredients effects of the tensor force are least
known so far. For instance, in most of the Relativistic Mean-Field
(RMF) models, the $E_{sym}(\rho)$ are determined by the coupling
schemes and properties of the $\rho$ and $\delta$ mesons.
Generally, no tensor coupling is considered. In phenomenological
models, such as the Skyrme and/or Gogny Hartree-Fock approaches,
the spin-isospin dependence of the three-body force is the most
uncertain term determining the density dependence of the
$E_{sym}(\rho)$ while effects of the tensor forces are normally
not considered either. On the other hand, most of the more
microscopic many-body theories using modern nucleon-nucleon
interactions have incorporated all major ingredients affecting the
$E_{sym}(\rho)$ albeit at different levels. Because of the
different many-body approaches and interactions used, although all these models are
well established and transparent, it has been hard to identify the main
causes for their different predictions for the $E_{sym}(\rho)$. To our
best knowledge, currently there is no community consensus
regarding the underlying physics responsible for the uncertain density dependence of
nuclear symmetry energy especially at supra-saturation densities.

Why is the density dependence of nuclear symmetry energy so
uncertain especially at supra-saturation densities? What are the effects of the tensor force?
To help answer these questions, using simple and phenomenological
approaches \cite{Xu11,Li11}, we have recently investigated
effects of the tensor force on the kinetic ($E_{sym}^{kin}(\rho)$)
and potential ($E_{sym}^{pot}(\rho)$) parts of the symmetry
energy, separately. The most striking finding is that, unlike the free Fermi gas model prediction
$E_{sym}^{kin}(\textrm{FG})(\rho)\equiv(2^{\frac{2}{3}}-1)(\frac{3}{5} \frac{\hbar^2 k_F^2}{2m})\approx 12.5\rho^{2/3}$
that has been widely used in both nuclear physics and
astrophysics, the tensor force induced high momentum tail in the
single-nucleon momentum distribution in SNM reduces significantly
the $E_{sym}^{kin}(\rho)$ to values much small than the $E_{sym}^{kin}(\textrm{FG})(\rho)$.
In fact, the $E_{sym}^{kin}(\rho)$ can become zero or
even negative if more than about $15\%$ nucleons populate the high-momentum
tail above the Fermi surface as indicated by the recent experiments done at
the Jefferson National Laboratory (J-Lab) by the CLAS
Collaboration \cite{CLAS}. It is very encouraging to note that not only
this finding was very recently confirmed qualitatively by three
independent studies using the state-of-the-art microscopic
many-body theories \cite{vid,carb,lov,vid12}, our calculation of the direct but second-order
tensor contribution to the $E_{sym}^{pot}(\rho)$ was also qualitatively verified by a more
accurate calculation very recently \cite{Wang12}.

\section{Tensor force induced short-range nucleon-nucleon correlation and its effect on single-nucleon momentum distribution in symmetric nuclear matter}

Our work was largely stimulated by the recent progress in
studying the tensor force induced nucleon short-range correlation
(SRC) and its consequence in single-nucleon momentum distribution
in SNM. Microscopic many-body theories indicate
that the tensor force affects both the single-nucleon momentum
distribution $n(k)$ and the two nucleon momentum distribution (as a function
of their total and relative momenta). Both have been
extensively studied theoretically and experimentally. For reviews,
see, e.g., refs. \cite{Bethe71,fra,Arr11}. Firstly, it is well
known that both the short-range repulsive core and the tensor
force acting in the isosinglet neutron-proton interaction channel lead to
SRC. Consequently, some nucleons are expected to be expelled from
below to above the Fermi surface leading to a high momentum tail
in the single-nucleon momentum distribution. Moreover, since the
hard core is approximately the same for all nucleon pairs while
the tensor force only acts between isosinglet neutron-proton
pairs, the difference in single-nucleon momentum distributions in
SNM and pure neutron matter (PNM) is mainly due to the tensor force
induced SRC. Since the isospin-dependence of SRC caused by the tensor force will
affect differently the EOSs of PNM and SNM especially at high densities,
the tensor force is expected to play an important role in
determining the $E_{sym}(\rho)$ which can be written as the difference
between the energy per nucleon in PNM and SNM, i.e., $E_{sym}(\rho)=E_{PNM}(\rho)-E_{SNM}(\rho)$, within the
parabolic approximation of the EOS of isospin-asymmetric nuclear matter.
On the other hand, effects of the short-range repulsive core which is essentially isospin-independent
on the $E_{sym}(\rho)$ largely cancel out. We noticed that because the local density
of SRC pairs in nuclei is estimated to reach that expected in the core of neutron stars,
it has been repeatedly speculated that the observed isospin-dependence of the
SRC may have significant effects on the EOS of cold dense neutron-rich nucleonic matter
and thus properties of neutron stars \cite{pia,sub}.

How does the SRC affect the single-nucleon momentum distribution $n(k)$?
This question has been studied both experimentally and
theoretically for a long time. In fact, many interesting
results have already been well established. For a comprehensive review, we refer
the reader to the book by Antonov et al. \cite{ant}. Probably the
most striking theoretical prediction that was later experimentally
verified is the universal high momentum tail in $n(k)$, i.e., the
high momentum part of $n(k)$ is independent of the mass number A
for finite nuclei and it is almost the same as for
infinitely large SNM \cite{fan,Pie92,cio}. This feature clearly
indicates the short-range origin of the high momentum tail.
Moreover, microscopic many-body calculations have shown that the underlying tensor force is
responsible for the universal high momentum tail between about
300-600 MeV/c \cite{Pie92}. At even higher momentum a weak
three-body SRC will show up. Successful efforts have been made to
extract experimentally from the scaling of inclusive electron
scattering cross sections on different targets the absolute
per-nucleon probability for nucleons to be in the high momentum
tail. In particular, an analysis of the experiments done by the CLAS Collaboration
\cite{CLAS} have shown quantitatively that the absolute values of the per-nucleon
probability of two-nucleon SRCs due to the tensor force is about
$15.4\pm 3.3\%,  19.3\pm 4.1\%, 22.7\pm 4.7\%$ for $^4$He, $^{12}$C
and $^{56}$Fe, respectively. Theoretically, to investigate how the strength of
the tensor force affects the high-momentum tail of $n(k)$, some
phenomenological methods have been particularly useful \cite{ant}.
For instance, Dellagiacoma \textit{et. al} have
derived formulas for $n(k)$ explicitly including the tensor force
induced SRC for finite nuclei \cite{dell,tra}. It was shown that
the probability of finding nucleons above the Fermi surface increases
with the strength of the tensor force (or equivalently the percentage
of D-wave mixture) \cite{dell,tra}.

Another kind of experiments especially useful for exploring the
SRC is the exclusive measurement of two-nucleon knockout reactions
induced by a high energy proton or electron. Recently, the isospin-dependence of the
SRC has been studied extensively both experimentally
\cite{pia,sub,Tan03,Bag10} and theoretically
\cite{alv,sch,nef,Dic04,Frick05,Sar05}. For reviews, see, e.g.,
Refs.\ \cite{fra,Arr11}. The experimental finding that the $np$
SRC dominates over the $nn~(pp)$ one indicates clearly that the
tensor force instead of the repulsive core is mainly responsible
for the high-momentum tail of $n(k)$. It is particularly exciting
to note that experiments at J-Lab have shown that about 20\% of
nucleons in $^{12}$C are correlated. This is quantitatively
consistent with the finding by the CLAS Collaboration from
inclusive electron scattering experiments. Moreover, the strength
of the $np$ SRC is found to be about 20 times that of the $pp~(nn)$ SRC
\cite{sub}. This has been shown as a direct consequence of the
tensor force acting in the deuteron-like $np$ state in the targets \cite{pia,alv,sch,nef}.

\begin{figure}[htb]
\centering
\includegraphics[width=10cm]{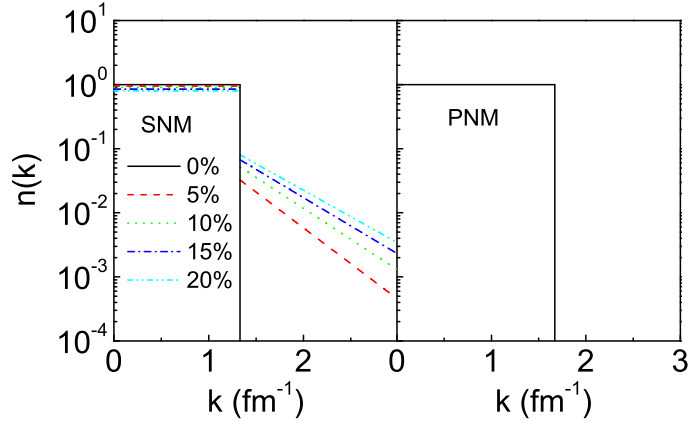}
\caption{Single-nucleon momentum distribution in symmetric nuclear
matter and pure neutron matter at normal density.} \label{n(k)}
\end{figure}

As mentioned above, one of our main goals is to understand at
least qualitatively effects of the tensor force induced SRC on the $E_{sym}(\rho)$.
Within a phenomenological model we can take into account the main features
of the $n(k)$ predicted by the microscopic many-body theories and confirmed by the experiments.
To explore effects of the tensor force, we can also easily vary the
percentage of nucleons in the high momentum tail and examine its
effects on the $E_{sym}(\rho)$. More specifically, we parameterize the $n(k)$ of nucleons in SNM as
\begin{eqnarray}\label{n(k)}
n(k) & = & a  \,\,\, (k \leq k_F) \\ \nonumber
     & = & e^{b\,k} \,\,\, (k > k_F).
\end{eqnarray}
The $a$ and $b$ are parameters determined by the normalization
condition $\frac{3}{k_F^3}\int_0^{\infty} n(k) k^2dk =1$ and the
percentage $\theta_{k\leq k_F}$ ($\theta_{k>k_F}$) of nucleons
below (above) the Fermi surface, i.e.,
\begin{eqnarray}
\frac{3}{k_F^3}\int_0^{k_F} n(k) k^2dk \times 100\% =
\theta_{k\leq k_F},\,\,\, \frac{3}{k_F^3}\int_{k_F}^{\infty} n(k)
k^2dk \times 100\% = \theta_{k> k_F}.
\end{eqnarray}
For PNM, the SRC is induced only by the repulsive core. The relatively weak n-n SRC indicated by the J-Lab experiments
justifies the use of an ideal gas approximation for the $n(k)$ in PNM.  Shown in the left panel of Fig.\ref{n(k)}
is the $n(k)$ in SNM at the saturation density for the ideal Fermi gas and $\theta_{k>k_F}=5\%$, $10\%$, $15\%$, \textrm{and} $20\%$, respectively.
For a comparison, the $n(k)$ for PNM is shown in the right panel.
It should be noted that the $n(k)$ for SNM with $\theta_{k>k_F}=20\%$ is very close to the
microscopic single-nucleon momentum distribution given by Ciofi
degli Atti et.al. \cite{cio} obtained by fitting results of the variational many-body calculations \cite{fan}.

\section{Tensor force effects on nuclear symmetry energy}
It has long been known that the tensor force influences the high density behavior of
$E_{sym}(\rho)$ \cite{Pan72,Wir88a,bro1,Eng98,xuli,Lee1}. However,
predictions from different models diverge widely partially because the strength of the in-medium tensor force and
its effects on the SRC were not so clearly known. The newly
available and more quantitative information on the tensor force
induced SRC may allow us to better understand effects of the
tensor force on the $E_{sym}(\rho)$ of dense neutron-rich nucleonic matter.
In the following, we explore separately effects of the tensor force on the kinetic and potential parts of the
symmetry energy.

\subsection{Tensor force effects on the kinetic part of nuclear symmetry energy}
\begin{figure}[htb]
\centering
\includegraphics[width=15cm]{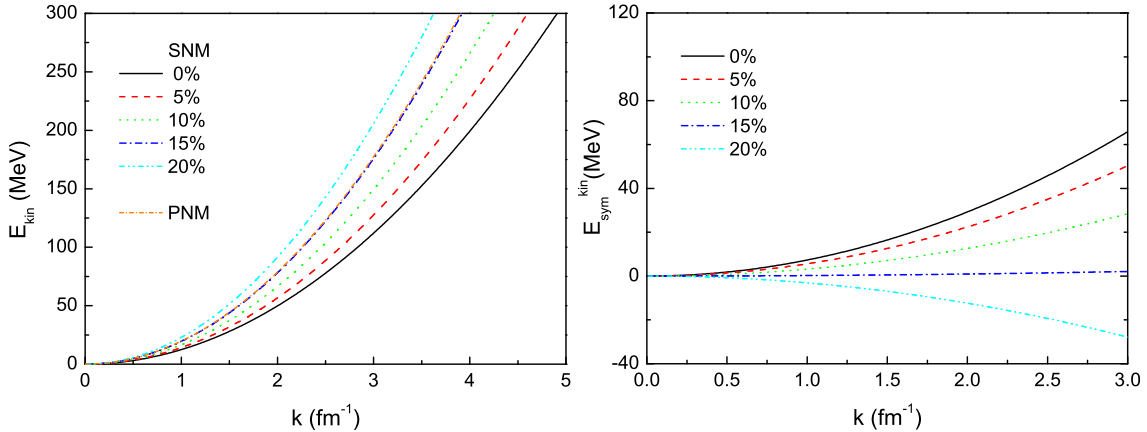}
\caption{Left window: The average kinetic energy per nucleon
$E_{kin}$ for pure neutron matter and symmetric nuclear matter
with different percentages of correlated nucleons
($\theta_{k>k_F}$) as a function of Fermi momentum. Right window:
The kinetic energy part of nuclear symmetry energy with different
percentages of correlated nucleons ($\theta_{k>k_F}$) as a
function of Fermi momentum.} \label{E_{kin}}
\end{figure}
While the average kinetic energy per nucleon in a Fermi gas of independent nucleons is simply
$E_{kin}=\frac{3}{5} \frac{\hbar^2 k_F^2}{2m}$, with correlated
nucleons it is given by
\begin{eqnarray}
E_{kin} = \alpha\int_0^{\infty} \frac{\hbar^2 k^2}{2m} n(k) k^2 dk,
\end{eqnarray}
where $\alpha=\frac{3}{k_F^3}$ and $\frac{3}{2k_F^3}$ for SNM and
PNM, respectively. We compare in the left window of Fig.\ref{E_{kin}}
the average kinetic energy as a function of Fermi momentum for PNM
and SNM with the percentage of high momentum nucleons to be
$\theta_{k>k_F}=0\%$, $5\%$, $10\%$, $15\%$, \textrm{and} $20\%$,
respectively. As one expects, the SRC increases the $E_{kin}$
significantly for SNM. More quantitatively, for SNM at the
saturation density corresponding to $K_F=$1.33 fm$^{-1}$, the
$E_{kin}$ with $\theta_{k>k_F}=20\%$ ($E_{kin}(k_F)\simeq40$ MeV)
is about twice of that ($E_{kin}(k_F)\simeq22$ MeV) for the free
Fermi gas. However, the $E_{kin}$ for PNM is the same as for the
free Fermi gas. Consequently, the tensor force induced high
momentum tail in SNM will affect the kinetic part of the nuclear
symmetry energy. In particular, if about $15\%$ nucleons in SNM
are in the high momentum tail, it is seen that the average kinetic
energy is about the same in PNM and SNM. This will then lead to an
approximately zero kinetic symmetry energy.

Shown in the right window of Fig.\ref{E_{kin}} is the kinetic part of
the symmetry energy as a function of Fermi momentum with different
$\theta_{k>k_F}$. Indeed, it is interesting to see that the tensor
force induced SRC has a significant impact on the kinetic part of
the symmetry energy, especially at supra-saturation densities. For
instance, the kinetic contribution to the symmetry energy
(E$_{sym}^{kin}(\rho)$) is negligibly small when 15\% nucleons are
considered to be correlated ($\theta_{k>k_F}=15\%$). With
$\theta_{k>k_F}=20\%$ when the tensor force is even stronger, the
$E_{sym}^{kin}(\rho)$ becomes negative at supra-saturation
densities. We are encouraged that this new features first observed
in our preliminary study \cite{Xu11} is also seen in very recent
studies based on the Brueckner-Hartree-Fock approach (BHF)
\cite{vid}, the Self-Consistent Green's Functions approach (SCGF)
\cite{carb}, and the Fermi-Hypernetted-Chain calculations (FHNC)
\cite{lov}. While all these models were long established, the
kinetic and potential contributions to the symmetry energy were
always combined together in previous studies. A careful
examination of their respective contributions was found very
informative \cite{vid,carb,lov}. Since the well-known and widely
used Fermi gas model prediction for the kinetic contribution
$E_{sym}^{kin}(\textrm{FG})(\rho)$ is always positive and
increases with increasing density, the dramatic tensor force
effects demonstrated by both the phenomenological and microscopic
models are conceptually important and practically useful.

At this point, some discussions regarding the implications of the
above results are in order. In many studies in both nuclear
physics and astrophysics, it is customary to write the total
symmetry energy as $E_{sym}(\rho)=12.5
(\rho/\rho_0)^{2/3}+E^{pot}_{sym}(\rho)$ where the first term is
the Fermi gas prediction for the $E_{sym}^{kin}(\rho)$ and the
$E^{pot}_{sym}(\rho)$ is the potential contribution. In doing so, however, one neglects completely
effects of the tensor force on the $E^{kin}_{sym}(\rho)$. As shown
above, the latter is approximately zero at normal density if
indeed about $15\%$ nucleons in SNM are above the Fermi surface as
indicated by the analysis of experiments at the J-Lab. Then, noticing
that the total symmetry energy at normal density is known to be about
30 MeV from analyzing the atomic masses and many other experiments,
if one believes in the $E^{pot}_{sym}(\rho)$ extracted from nuclear
reactions and the almost zero $E^{kin}_{sym}(\rho)$ at normal density,
an interesting question arises: where is the remaining symmetry
energy? As we shall show next, it is in the tensor contribution to the potential part of
the symmetry energy.

\subsection{Second-order tensor force contribution to the potential part of the symmetry energy}
It is well known that the first-order (at the mean-field level) tensor contribution to the EOS
vanishes in spin-saturated systems. In the best-studied phenomenology of nuclear forces, i.e., the
one-boson-exchange model, the tensor interaction results from the
exchange of an isovector $\pi$ and/or $\rho$ meson. For instance,
the tensor part of the one-pion exchange potential (OPEP) can be
written in configuration space as~\cite{mac89}
\begin{equation}
 V_{t\pi}= -\frac{f_{\pi}^2}{4\pi}m_{\pi}(\tau_1\cdot\tau_2)S_{12}
[\frac{1}{(m_{\pi}r)^3}+\frac{1}{(m_{\pi}r)^2}+\frac{1}{3m_{\pi}r}]\exp(-m_{\pi}r)
\label{pi}
\end{equation}
where $r$ is the interparticle distance and
$S_{12}=3\frac{(\sigma_1 \cdot r)(\sigma_1 \cdot
r)}{r^2}-(\sigma_2\cdot\sigma_2)$ is the tensor operator. The
$\rho$-exchange tensor interaction $V_{t\rho}$ has the same
functional form as the OPEP, but with the $m_{\pi}$ replaced
everywhere by $m_\rho$, and the $f_{\pi}^2$ by $-f_{\rho}^2$. The
magnitudes of both the $\pi$ and $\rho$ contributions grow quickly
with decreasing $r$. A proper cancellation of these opposite
contributions leads to a realistic strength for the nuclear tensor
force. However, since the tensor coupling is not well determined
consistently from deuteron properties and/or nucleon-nucleon
scattering data, the tensor interaction is by far the most
uncertain part of the nucleon-nucleon interaction~\cite{mac01}. In
addition, due to both the physical and mathematical differences in
methods used during construction~\cite{mac01}, various realistic
nuclear potentials usually have widely different tensor components
at short range ($r\leq$0.8 fm). For example, in the Paris
potential~\cite{Paris}, it is just described simply by a constant
soft core. The Argonne V18 (AV18) uses local functions of
Woods-Saxon type~\cite{av18}, while Reid93 applies local Yukawas
of multiples of the pion mass ~\cite{Reid93}. While it is
promising that new experiments, such as, (p,d) reactions induced
by high energy protons~\cite{tan10} or two-nucleon knockout
reactions induced by high energy electrons~\cite{sub,Bag10}, may
allow us to better constrain the short-range tensor force in the
near future, currently the short-range behavior of the tensor
force is still very uncertain. Here we use several typical and
widely used tensor forces that are the same at long-range as the
one used in the AV18 but have characteristically different
short-range behaviors. This will allow us to examine effects of
the short-range tensor force on the density-dependence of nuclear
symmetry energy.

It is easy to see from Eq.~\ref{pi} that the expectation value of
the tensor force $<V_t>$ is zero in spin-saturated systems. Thus, the first-order tensor force
does not contribute to the symmetry energy unless one assumes that
all isosinglet neutron-proton pairs behave as bound deuterons~\cite{Xuli10a}.
Thus, it is the second-order tensor contribution that is important for the binding energy of nuclear
matter~\cite{kuo65,bro81} and also for the symmetry
energy~\cite{mac94}. Using a second-order effective tensor
interaction obtained first by Kuo and Brown~\cite{kuo65}, see. e.g.,
ref.~\cite{sob} for a review, the tensor contribution to the
symmetry energy is approximately ~\cite{mac94}
\begin{equation}\label{Mac}
<V_{sym}> = \frac{12}{e_{\rm eff}}<V_t^2(r)>
\end{equation}
where the $e_{\rm eff}\approx 200$ MeV around normal density.
While this approximate expression may lead to symmetry energies
systematically different from predictions of advanced microscopic
many-body theories using various interactions especially at high densities, it is handy to
evaluate effects of the different short-range tensor forces within
the same analytical approach. To evaluate the expectation value of
$V_{\texttt{{\rm sym}}}$, we use the free single-particle wave
function $(V^{-1}e^{i \textbf{k}\cdot \textbf{r}})\eta_{
\lambda}\zeta_{\tau}$, where $\eta_{\lambda=\uparrow/\downarrow}$
and $\zeta_{\tau=p/n}$ is the spin and isospin wave function,
respectively. The direct and exchange matrixes are, respectively,
\begin{eqnarray}
&& \langle~ \textbf{k} \lambda \tau \textbf{k}' \lambda'
\tau'|V_{{\rm sym}}|\textbf{k} \lambda \tau \textbf{k}' \lambda'
\tau'\rangle
= \frac{1}{V} \int
V_{{\rm sym}}(\textbf{r})d^3 r \\
{\rm and} \nonumber \\
&& \langle \textbf{k} \lambda \tau \textbf{k}' \lambda'
\tau'|V_{{\rm sym}}|\textbf{k}' \lambda' \tau' \textbf{k} \lambda
\tau \rangle
= \frac{1}{V} \delta_{\lambda\lambda'}\delta_{\tau\tau'} \int
\exp[-i(\textbf{k}-\textbf{k}')\cdot \textbf{r}] V_{{\rm
sym}}(\textbf{r})~d^3 r.
\end{eqnarray}
The expectation value of $V_{{\rm sym}}$ in the $S = 1, T = 0$
channel is thus
\begin{eqnarray}
&& <V_{{\rm sym}}> =  \frac{1}{16} \frac{1}{2} \sum_{\textbf{k}
\lambda \tau} \sum_{\textbf{k}' \lambda' \tau'} [\langle
\textbf{k} \lambda \tau \textbf{k}' \lambda' \tau'|V_{{\rm
sym}}|\textbf{k} \lambda \tau \textbf{k}' \lambda' \tau'\rangle -
\langle \textbf{k} \lambda \tau \textbf{k}' \lambda' \tau'|V_{{\rm
sym}}|\textbf{k}' \lambda' \tau' \textbf{k} \lambda \tau \rangle]
\nonumber
\\ &&=\frac{V}{2}\frac{1}{(2\pi)^6}\int^{k_F}d^3k\int^{k_F}d^3k'
\{\int V_{{\rm sym}}(\textbf{r})d^3 r -\frac{1}{4}\int
\exp[-i(\textbf{k}-\textbf{k}')\cdot \textbf{r}] V_{{\rm
sym}}(\textbf{r})~d^3 r\}.
 \label{eq:vsymm}
\end{eqnarray}
Noticing that the momentum integral $\int^{k_F}d^3k e^{i
\textbf{k}\cdot \textbf{r}}= 4\pi \int_0^{k_F} k^2 j_0(kr)dk=
\frac{4\pi k_F^3}{3} \frac{3j_1(k_Fr)}{k_Fr}$ and the particle
number density $\frac{A}{V}=\frac{2}{3\pi^2}k_F^3$, we can write
the tensor contribution to the symmetry energy as
\begin{equation}
\frac{<V_{{\rm sym}}>}{A} = \frac{12}{e_{\rm
eff}}\cdot\frac{k_F^3}{12\pi^2} \{\frac{1}{4}\int V_{t}^2(r)d^3 r
- \frac{1}{16}
\int[\frac{3j_1(k_Fr)}{k_Fr}]^2V_{t}^2(r)d^3 r\}.
\end{equation}
For large $k_F$, the second integral in the above equation
approaches zero, the first term is thus expected to dominate at
high densities, leading to an almost linear density dependence.

\begin{figure}
\centering
\resizebox{0.65\textwidth}{!}{%
\includegraphics{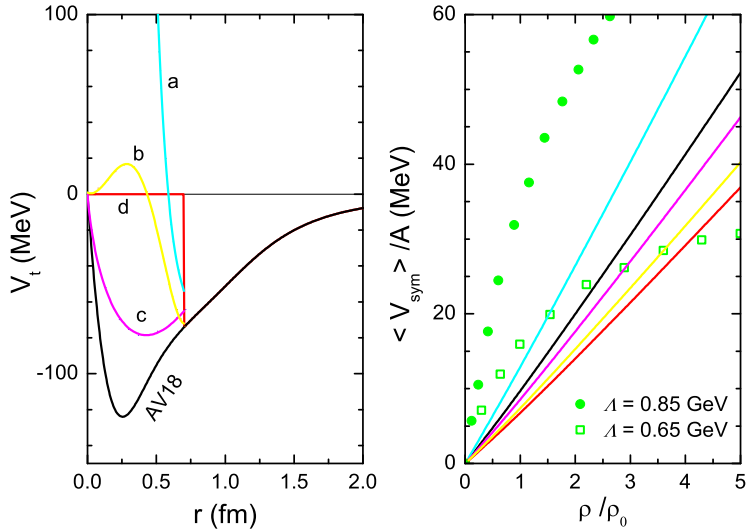}
} \caption{Left panel: different short-range tensor
interactions with the same AV18 long-range part for $r > 0.7$fm;
Right panel: potential part of the symmetry energy with the
different short-range tensor interactions. The (filled) open
(circles) squares are the 2-particle-2-hole calculations from ref.
\protect{\cite{Wang12}}.} \label{f1}
\end{figure}

To access quantitatively effects of the short-range tensor force
on the density dependence of nuclear symmetry energy, we adopt
here several tensor forces used by Otsuka et al. in their recent
studies of nuclear structures~\cite{ots05}. The considered tensor
forces, including the standard $\pi+\rho$ exchange (labelled as
$a$), the G-Matrix (GM)~\cite{ots05} (labelled as $b$),
M3Y~\cite{m3y}(labelled as $c$), and the AV18~\cite{av18}
(labelled as AV18), as shown in the left panel of Fig.~\ref{f1},
behave rather differently at short distance, but merge to the same
AV18 tensor force at longer range. In addition, we add a case
($d$) where the tensor force vanishes for $r\leq 0.7$ fm. The
$\pi+\rho$ exchange interaction is fixed by the standard
meson-nucleon coupling constants with a strong $\rho$
coupling~\cite{sob}, and we use a short-range cut-off at $r=0.4$
fm, i.e.,  $V(r<0.4{\rm~fm})= V(r=0.4{\rm~fm})$. The resulting
tensor contribution to the nuclear symmetry energy is shown as a
function of density in the right panel of Fig.~\ref{f1}. As
expected, they tend to grow linearly with increasing density.
Since it is the square of the tensor force that determines its
contribution to the symmetry energy, tensor forces having larger
magnitudes at short distance affect more significantly the
symmetry energy. It is seen that the variation of the tensor force
at short distance affects significantly the high-density behavior
of nuclear symmetry energy.

Around normal density where predictions based on Eq.~\ref{Mac} are
most reliable, the tensor force contribution to the symmetry
energy is about 7 to 15 MeV depending the interaction used.
Generally speaking, this can largely compensate the tensor force
induced reduction in the kinetic part of the symmetry energy.
While in some microscopic models, the tensor force induced effects
may have been treated self-consistently in calculating both the
kinetic and potential parts of the symmetry energy, in almost all
phenomenological models they are not considered at all. However,
they can all be adjusted to give the correct symmetry energy at
least at normal density. Our analysis here indicates that this is
not surprising largely because of the approximate cancelation
between the tensor force induced reduction of the
$E^{kin}_{sym}(\rho)$ and its second-order contribution to the
$E^{pot}_{sym}(\rho)$.

In a recent study by Wang et al. \cite{Wang12}, the tensor force
contribution to the potential part of the symmetry energy due to the one-pion exchange
was evaluated accurately at the 2-particle-2-hole level. While their results depend on the momentum
cut-off parameter $\Lambda$ in the form factor, they demonstrated clearly
that the potential part of the symmetry energy has a large tensor
contribution. For a comparison, their results with $\Lambda=0.85$
GeV and $\Lambda=0.65$ GeV which are within the range of theoretical
expectations for this parameter \cite{Wang12,Kai02} are also shown in Fig. \ref{f1}. Qualitatively, their results are consistent with our results
based on Eq.~\ref{Mac} in the sense that the direct, second-order tensor contribution to the potential part of the $E_{sym}(\rho)$ is large. However, as most other calculations in the literature, effects of the tensor force on the kinetic part of the symmetry energy was not considered in
their work either. With the strong tensor force they used, the kinetic part of the symmetry energy is likely to become
zero or negative as we discussed in the previous subsection. To this end, it is also worth noticing that in essentially all mean-field based
models without considering high-order tensor contributions, the interaction parameters are normally adjusted to give the $E_{sym}(\rho_0)$ at normal density a value of about 30 MeV. The discussions above indicate that these parameters need to be
re-adjusted if the tensor force contributions in either or both kinetic and potential terms are also considered.

\subsection{Tensor force effects on the central force contribution to the symmetry energy}
\begin{figure}[htb]
\centering
\includegraphics[width=10cm]{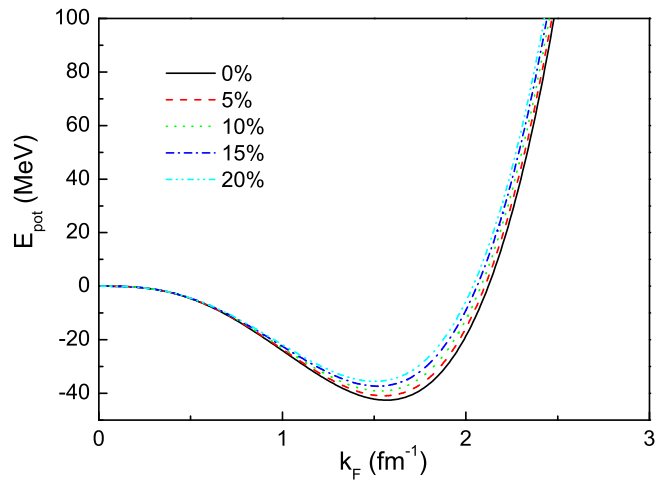}
\caption{The average potential energy per nucleon $E_{pot}$ for
symmetric nuclear matter with different percentages of correlated
nucleons ($\theta_{k>k_F}$) as a function of Fermi momentum.} \label{E_{pot}}
\end{figure}

Because of the spin-isospin dependence of the central force, it
also contributes to the symmetry energy. Moreover, if finite-range
interactions are considered, the corresponding single-particle
potential is momentum dependent and the potential energy density
involves a double integration over the momenta of two interacting
nucleons. Thus, the tensor force induced high-momentum tail may
also affect the central force contribution to the EOS of SNM. The
central force contribution to the symmetry energy may then also be
affected. To our best knowledge, this effect was previously
ignored in all phenomenological models. To evaluate this effect,
we use here the MDI (Momentum-Dependent Interaction)\cite{Das03}
potential which has been used extensively in transport models
simulations of heavy-ion reactions \cite{BALi04}. The MDI
potential energy per nucleon can be written as (with the parameter $x=0$)\cite{Das03}
\begin{equation}
E_{pot} = \frac{A}{2} \frac{\rho}{\rho_0} + \frac{B}{\sigma+1}
\frac{\rho^{\sigma}}{\rho_0^{\sigma}} + \frac{C}{\rho\rho_0}
\int_0^{\infty} \int_0^{\infty}
\frac{n(k_1)n(k_2)}{1+(\vec{k_1}-\vec{k_2})^2/\Lambda_k^2}d\vec{k_1}d\vec{k_2},
\end{equation}
where $\sigma=4/3$ and the parameter $\Lambda_k=1.0 k_F^0$
\cite{Das03}. The $n(k_1)$ and $n(k_2)$ are the one-body momentum
distribution of nucleon-1 and nucleon-2, respectively. The
parameters A, B and C were determined by fitting the empirical
properties of SNM without considering effects of the tensor force
induced SRC as the MDI potential was based on a mean-field model
(i.e., the Hartree-Fock with a modified Gogny force). In
principle, with each tensor force leading to a different high
momentum tail in SNM, one has to readjust all associated
parameters to reproduce the empirical properties of SNM. However,
for the purpose of this study it is actually advantageous to keep
using the same original parameters so that effects of the SRC can
be clearly revealed. Thus, in the present calculations, we do not
modify the original parameters of the MDI interaction density
functional.

\begin{figure}[htb]
\centering
\includegraphics[width=14cm]{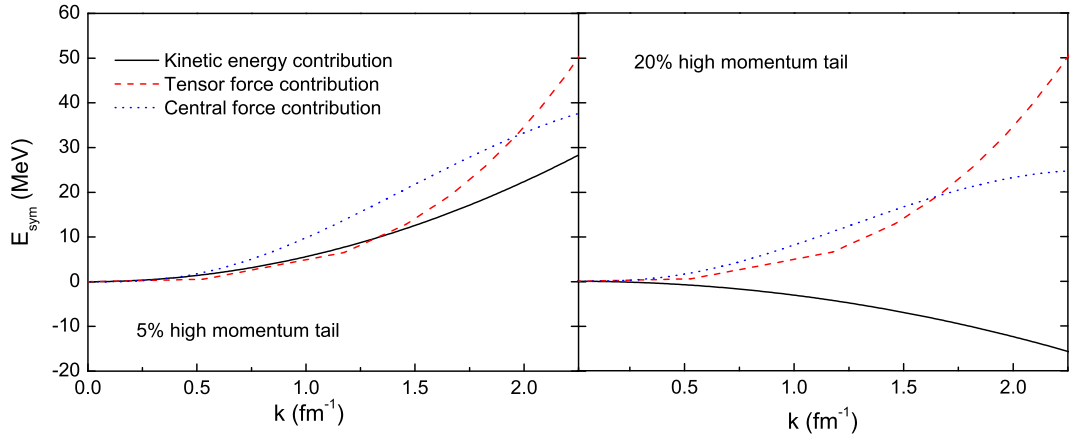}
\caption{The kinetic, central force and tensor force contributions to the symmetry energy with $5\%$ and
$20\%$ high-momentum nucleons as a function of Fermi momentum.} \label{E_{sym}}
\end{figure}

Shown in Fig.\ref{E_{pot}} is a comparison of $E_{pot}$ with
different percentages $\theta_{k>k_F}$ of correlated nucleons for
SNM. Similar to the case of the average kinetic energy $E_{kin}$,
the SRC increases the potential energy $E_{pot}$ for SNM as one
expects. Since the tensor force has no effect on the EOS of PNM,
an increase in $\theta_{k>k_F}$ will lead to a decrease in central
force contribution to the symmetry energy. This is clearly seen in
Fig.\ref{E_{sym}} by comparing the central force contributions to
the symmetry energy with $\theta_{k>k_F}=5\%$ and $20\%$ given in
the left and right window, respectively. It is also interesting to
compare relative effects of the tensor force on the symmetry
energy through the kinetic, central force and the direct
second-order potential term. Obviously, the latter is the
dominating mechanism for the tensor force to affect the symmetry
energy. The kinetic energy comes second and the central force
contribution is least affected by the tensor force. As to their
relative contributions to the total symmetry energy, it is seen
that they are all comparable and important to the total symmetry
energy.

\section{Summary}
Using phenomenological models, we explored effects of the tensor force on the density dependence of nuclear symmetry energy.
The high momentum tail in symmetric nuclear matter induced by the tensor force acting between protons and neutron makes the kinetic part of the symmetry energy $E^{kin}_{sym}(\rho)$ significantly smaller than the Fermi gas model prediction. With about $15\%$ nucleons in the high momentum tail in SNM as indicated by the recent experiments at the J-Lab, the $E^{kin}_{sym}(\rho)$ is negligibly small. It even  becomes negative when more nucleons are in the high momentum tail in SNM. While at the mean-field level the tensor force
has no contribution to the EOS, its second-order contribution to the
potential part of the symmetry energy is large. To completely take
into account effects of the tensor force, it is necessary to include not only its second-order
potential contribution and effects on the kinetic part but also
its effects on the central force contribution to the symmetry
energy. Implications of these finding on extracting experimental
constraints on the density dependence of nuclear symmetry energy
are also discussed briefly.

\ack
We would like to thank L.W. Chen, A. Lovato, A. Rios, I. Vidana and W. G. Newton for
helpful discussions and communications. This work is supported in part by the US National Science Foundation grant
PHY-0757839 and PHY-1068022, the National Aeronautics and Space
Administration under grant NNX11AC41G issued through the Science
Mission Directorate, the National Natural Science Foundation of
China (Grants 10805026, 10905048 and 11175085) and the National
Basic Research Program of China under Grant 2009CB824800.

\end{document}